\documentclass[aps,preprint,epsfig,rotate]{revtex4}
\begin{document}
\title{Accurate evaluations of the field shift and lowest-order QED correction for the ground $1^1S-$states of some light 
       two-electron ions.}

 \author{Alexei M. Frolov}
 \email[E--mail address: ]{afrolov@uwo.ca}

\affiliation{Department of Applied Mathematics \\
 University of Western Ontario, London, Ontario N6H 5B7, Canada}

\author{David M. Wardlaw}
 \email[E--mail address: ]{dwardlaw@mun.ca}

\affiliation{Department of Chemistry, Memorial University of Newfoundland, St.John's, 
             Newfoundland and Labrador, A1C 5S7, Canada}

\date{\today}

\begin{abstract}
Mass-dependent and field shift components of the isotopic shift are determined to high accuracy for the ground $1^1S-$states of some 
light two-electron Li$^{+}$, Be$^{2+}$, B$^{3+}$ and C$^{4+}$ ions. To determine the field components of these isotopic shifts we 
apply the Racah-Rosental-Breit formula. We also determine the lowest order QED corrections to the isotopic shifts for each of these 
two-electron ions.

\end{abstract}

\maketitle
\newpage

\section{Introduction}

In this study we perform highly accurate computations of the isotopic shifts for the ground $1^1S-$states of some light two-electron ions: 
Li$^{+}$, Be$^{2+}$, B$^{3+}$ and C$^{4+}$. In atomic and molecular spectroscopy the isotopic shift \cite{Sob}, \cite{LLQ} usually means 
the difference (or shift) in the total and/or binding energies of the bound states that occurs when one nuclear isotope is replaced by 
another. It is clear $a$ $priori$ that the total energies and other bound state properties of light atoms and ions depend upon the inverse 
mass of the central nucleus and proton density distribution in that nucleus. In some cases a few other nuclear properties, e.g., the nuclear 
magnetic moment, also contribute to the total energies of atoms and ions, and therefore, to the isotopic shifts. In this study we perform 
highly accurate evaluations of the different components of isotopic shifts in the light two-electron ions Li$^{+}$, Be$^{2+}$, B$^{3+}$ and 
C$^{4+}$. Our approach is essentially non-relativistic, i.e. we use the non-relativistic wave functions which are determined as the solutions 
of the non-relativistic Schr\"{o}dinger equation for each of these two-electron ions. For heavier two-electron ions, e.g., for the O$^{6+}$, 
F$^{7+}$ and Ne$^{8+}$ ions, the contribution of the relativistic and QED corrections rapidly increases with the nuclear charge $Q e$ (or 
$Q$) and isotopic shifts for such ions can be determined to high accuracy only with the use of the relativistic bi-spinor wave functions 
which must be obtained from the Dirac equation(s). The wave functions arising from the non-relativistic Schr\"{o}dinger equation can be 
applied to such heavy two-electron ions to determine only approximate values of some lowest-order relativistic and QED-corrections.

In this study numerical evaluations of the isotopic shifts in light atoms and ions are based on highly accurate computations of the 
expectation values of a few selected electron-nuclear and electron-electron operators. At the first stage of our procedure we apply 
the wave functions which have been determined for model ions which have an infinitely heavy nucleus. By using these wave functions 
we determine the expectation values of operators which are included in different components of the isotopic shift. Formally, these 
expectation values allow us to evaluate the isotopic shifts (in the lowest-order approximation and to relatively high accuracy) for the 
ground states of all two-electron ions considered in this study (i.e. in the Li$^{+}$, Be$^{2+}$, B$^{3+}$ and C$^{4+}$ ions). However, 
the overall accuracy of our evaluations can be drastically improved, if we determine the same expectation values and the total energies for 
the two-electron ions with finite-mass nuclei. Indeed, each atomic system includes a central atomic nucleus and the mass of such a nucleus
is always finite. An error in the total energy due to replacement of the actual, finite mass nucleus by an infinitely heavy nucleus can 
be evaluated as $\approx 1 \cdot 10^{-6} - 1 \cdot 10^{-5}$ $a.u.$ For other bound state properties which can also contribute to various 
corrections to the total energy such (relative) errors vary between $\approx 1 \cdot 10^{-7}$ and $\approx 1 \cdot 10^{-3}$. Furthermore, 
there are additional corrections related to the fact that all nuclear masses are known only approximately. In reality, this means that all 
nuclear masses are the subject to constant experimental revision. It is clear that to solve the problem of isotopic shift completely we need 
to determine the mass gradients for each of the expectation values used to evaluate the isotope shifts in the same two-electron ions with 
varying nuclear masses. 

This work has the following structure. Representation of the isotopic shift in atoms as the sum of its leading components is discussed in 
Section II. In that Section we also investigate the formula which is used to determine the field component of the isotopic shift. Calculations 
of isotopic shifts for some light two-electron ions are performed in Section III, namely for the ground $1^1S-$states in the Li$^{+}$, 
Be$^{2+}$, B$^{3+}$ and C$^{4+}$ ions with an infintely heavy central nucleus. Here we also discuss a system of tests for the non-relativistic 
wave functions which can be used in actual calculations. Section IV contains results of highly accurate computations for the two-electron ions 
with the finite nuclear masses. This Section is a central part of our study. Section V contains formulas for calculation of the lowest order 
QED correction in two-electron ions. Concluding remarks can be found in the last Section.   

\section{Components of the isotopic shift}

In general, the isotopic shift $\Delta E$ of the bound state level with the total energy $E$ can be represented as the sum of a 
few different components. In many cases the two largest components in such sums are: (a) the mass component $\Delta E_M$, which explicitly 
depends on the mass of the central nucleus, and (b) the field shift component $\Delta E_F$, which mainly depends upon the electric charge 
distribution in the atomic nucleus. The first component $\Delta E_M$ is represented as the sum of the normal and specific components. Each 
of these two components is proportional to the factor $\frac{m_e}{M}$, where $m_e$ is the mass of the electron at rest, while $M$ is the nuclear 
mass (at rest) expressed in $m_e$. For few-electron ($N-$electron) atoms and ions the exact formula for the isotopic shift $\Delta E_M$ takes the 
form (see, e.g., \cite{FrFi})
\begin{eqnarray}
 \Delta E_M = \Delta E^{nms}_{M} + \Delta E^{sms}_{M} = \frac{m_e}{M} \langle \sum^{N}_{i=1} 
 \frac{{\bf p}^2_i}{2 m_e} \rangle + \frac{m_e}{M} \langle \sum^{N}_{i (i \ge k) = 2} \sum^{N-1}_{k = 1}
 \frac{{\bf p}_i \cdot {\bf p}_k}{2 m_e} \rangle  \label{eq1}
\end{eqnarray}
where $\Delta E^{nms}_{M}$ is the normal mass shift, $\Delta E^{sms}_{M}$ is the specific mass shift and the notation $\langle \hat{X} 
\rangle$ designates the expectation value of the operator $\hat{X}$. For the two-electron (or helium-like) atoms and ions the 
expressions for the normal and specific components of the isotope shift are 
\begin{eqnarray}
 \Delta E^{nms}_{M} = \frac{1}{M} \langle {\bf p}^2_1 \rangle \; \; \; {\rm and} \; \; \; 
 \Delta E^{sms}_{M} = \frac{1}{2 M} \langle {\bf p}_1 \cdot {\bf p}_2 \rangle \; \; \; , 
 \label{eq3}
\end{eqnarray}
respectively. As follows from Eq.(\ref{eq3}), to determine the normal and specific components in two-electron atom/ion one needs to obtain 
the expectation values of the ${\bf p}^2_1$ and ${\bf p}_1 \cdot {\bf p}_2$ operators. Everywhere in this study we assume that the wave 
functions of the two-electron atom/ion are properly symmetrized upon spin-spatial permutations of the two electrons and, therefore, the 
corresponding single-electron expectation values are always equal to each other, e.g., $\langle {\bf p}^2_1 \rangle = \langle {\bf p}^2_2 
\rangle$.

In actual two-electron ions and atoms, i.e. in atomic systems with the finite nuclear mass $M$, one can use the condition which follows 
from the conservation of the total momentum ${\bf P}_N = {\bf p}_1 + {\bf p}_2$, where ${\bf P}_N$ is the momentum of the nucleus, while ${\bf 
p}_1$ and ${\bf p}_2$ are the electron momenta. From here one finds:   
\begin{eqnarray}
 \frac12 \langle {\bf P}^2_N \rangle = \langle {\bf p}^2_1 \rangle + 
 \langle {\bf p}_1 \cdot {\bf p}_2 \rangle  \label{eq4}
\end{eqnarray}
and, therefore, from Eq.(\ref{eq1}) for $N = 2$ and Eq.(\ref{eq4}) we obtain
\begin{eqnarray}
 \Delta E = \frac{1}{2 M} \langle {\bf P}^2_N \rangle \label{eq5}
\end{eqnarray}
i.e. the mass-dependent component of the isotopic shift is the expectation value of the kinetic energy of the atomic nucleus with the finite 
mass. In many books and textbooks the formula, Eq.(\ref{eq5}), is considered as the original (or fundamental) expression, while 
Eq.(\ref{eq1}) is derived from this formula.  

\subsection{The field component of the isotope shift}

In contrast with the mass component $\Delta E_M$, Eq.(\ref{eq5}), the field component of the isotopic shift $\Delta E_F$ explicitly depends upon 
the nuclear size (or nuclear radius) $R$ and proton density distribution in that nucleus. It is clear that this component also depends upon the 
nuclear mass $M$, since nuclear matter is a saturated matter (in contrast with Coulomb matter). The nuclear radius $R$ is uniformly related to the 
number of nucleons $A$ in the nucleus: $R = r_0 \cdot A^{\frac13}$, where the `constant' radius is $r_0 \approx 1.17 - 1.25 \cdot 10^{-13}$ $cm$ = 
1.17 - 1.25 $fm$ (fermi), where 1 $fm$ = $1 \cdot 10^{-13}$ $cm$. Briefly, this means that the field component of the total isotopic shift is also 
a function of the nuclear mass $M$, since $A \approx \frac{M}{m_p}$, where $m_p$ is the proton mass. In general, the nuclear mass is a function of 
$A, Z (= N_p)$, where $Z$ is the electric charge of the nucleus = number of protons $N_p$, and $N_n$ is number of neutrons. The formula for $M(A,Z)$ 
is known as the Weiz\"{a}cker formula. This formula is discussed in the Appendix. 

The field component of the isotopic shift $\Delta E_{F}$ is determined by the expression which is widely known as the Racah-Rosental-Breit formula 
(see, e.g., \cite{Sob} and references therein). In atomic units ($\hbar = 1, e = 1, m_e = 1$) this formula takes the form 
\begin{eqnarray}
 \Delta E_{F} = \frac{4 \pi a^2_0}{Q} \cdot \frac{b + 1}{[\Gamma(2 b + 1)]^2} \cdot B(b) \cdot \Bigl( \frac{2 Q R}{a_0} \Bigr)^{2 b} \cdot 
 \frac{\delta R}{R} \cdot \langle \delta({\bf r}_{eN}) \rangle \label{eqf3}
\end{eqnarray}
where $Q$ is the nuclear charge, $R$ is the nuclear radius and $b = \sqrt{1 - \alpha^2 Q^2}$, where $\alpha = \frac{e^2}{\hbar c} \approx 
\frac{1}{137}$ is the dimensionless constant which is the small parameter in QED. In Eq.(\ref{eqf3}) the notation $\Gamma(x)$ stands for the Euler's 
gamma-function, while the factor $B(b)$ is directly related to the proton density distribution in the atomic nucleus. By assuming a uniform 
distribution of the proton density over the volume of the nucleus one finds the following expression for the factor $B(b)$
\begin{eqnarray}
 B(b) = \frac{3}{(2 b + 1) (2 b + 3)} \label{eqf4} 
\end{eqnarray}
For light nuclei with $Q \le 6$ we have $b \approx 1$ and $B \approx \frac15$. The formula, Eq.(\ref{eqf3}), has been used in many papers for numerical 
evaluations of the field component of the isotopic shift, or field shift, for short. In some works, however, this formula was written with a number of 
`obvious simplifications'. Many such `simplifications' are based on the fact that for light nuclei the numerical value of the factor $b$ is close to 
unity. Furthermore, in some papers the factor $b$ was mistakenly called and considered as the Lorentz factor, while the actual Lorentz factor $\gamma$ 
is the inverse value of $b$, i.e., $\gamma = \frac{1}{b} = \frac{1}{\sqrt{1 - \alpha^2 Q^2}}$, which always exceeds unity. As follows from Eq.(\ref{eqf3})
in order to determine the field component of the isotopic shift in light atoms one needs to know the radius of the nucleus $R$ and the expectation value 
of the electron-nucleus delta-function $\langle \delta({\bf r}_{eN}) \rangle$.

In this study we evaluate the field components of the isotopic shift for a number of the ground $1^1S(L = 0)-$states in light two-electron ions by using 
the exact formula, Eq.(\ref{eqf3}). This allows one to evaluate the numerical errors which arise from the use of approximate expressions. As follows from 
Eq.(\ref{eqf3}), to evaluate the field component of the field shift one needs to determine to very high accuracy the expectation value of the electron-nuclear 
delta-function, i.e. $\langle \delta({\bf r}_{eN}) \rangle$. In the lowest-order approximation the ratio $\frac{\delta R}{R}$ in Eq.(\ref{eqf3}) can be 
assumed to be equal unity. The formula for $\Delta E_{F}$ is reduced to the form (in atomic units)
\begin{eqnarray}
 \Delta E_{F} = 4^{b+1} \pi Q^{2b-1} \cdot \alpha^{4 b} \cdot \frac{3 (b + 1)}{[\Gamma(2 b + 1)]^2 (2 b + 1) (2 b + 3)} \cdot \Bigl( \frac{R}{r_e} 
 \Bigr)^{2 b} \cdot \langle \delta({\bf r}_{eN}) \rangle \label{eqf5}
\end{eqnarray}
where $r_e = \alpha^2 a_0 \approx 2.817940$ $fm$ (1 $fm$ = $1 \cdot 10^{-13}$ $cm$ is one $fermi$) is the classical radius of the electron. For atomic nuclei 
the dimensionless factor $\frac{R}{r_e}$ in the last formula is close to unity. Also, in our calculations we have used the following numerical values for the 
physical constants: $\alpha = 7.2973525698 \cdot 10^{-3}$ and $a_0 = 5.2917721092 \cdot 10^{-9}$ $cm$. The formula, Eq.(\ref{eqf5}), has been used in all 
calculations of $\Delta E_F$ performed in this study. As follows from Eq.(\ref{eqf5}) to determine the field component of the isotopic shift one needs to know 
the expectation value of the electron-nuclear delta-function $\delta({\bf r}_{eN})$ and numerical value of the nuclear radius $R$. The expectation value $\langle 
\delta({\bf r}_{eN}) \rangle$ can be found from the results of highly accurate atomic computations, while the nuclear radii of different light nuclei must be 
taken from nuclear experiments (see, e.g., \cite{Angel}).

\section{Bound state calculations of the two-electron ions}

Our method used in this study to evaluate different components of the isotopic shift and the lowest-order QED corrections is based on numerical, highly accurate
computations of expectation values of some operators. In such calculations we apply the non-relativistic wave functions of the two-electron ions, which are 
obained as the solutions of the Schr\"{o}dinger equation \cite{FrFi} for the bound states $H \Psi = E \Psi$, where $E < 0$ and $H$ is the non-relativistic 
Hamiltonian of the two-electron ions 
\begin{equation}
 H = -\frac{\hbar^{2}}{2 m_{e}} \Bigl( \nabla^{2}_{1} + \nabla^{2}_{2} + \frac{m_e}{M_N} \nabla^{2}_{N} \Bigr) + \frac{Q e^2}{r_{32}} + \frac{Q e^2}{r_{31}} + 
 \frac{e^2}{r_{21}} \label{Ham}
\end{equation}
where $\nabla_{i} = \Bigl( \frac{\partial}{\partial x_{i}}, \frac{\partial}{\partial y_{i}}, \frac{\partial}{\partial z_{i}} \Bigr)$ and $i = 1, 2, 3(= N)$, where 
the notation $N (= 3)$ stands for the nucleus. In Eq.(\ref{Ham}) the notation $\hbar$ stands for the reduced Planck constant, i.e. $\hbar = \frac{h}{2 \pi}$, and 
$e$ is the elementary electric charge. Everywhere below in this study the particles 1 and 2 mean the electrons, while the particle 3 is the atomic nucleus with the 
mass $M_N \gg m_e$. The dimensionless ratio $\tau_{m} = \frac{m_e}{M_N}$ is the small parameter of the method. For light atoms it is very convenient to perform all 
bound state calculations in atomic units in which $\hbar = 1, m_e = 1$ and $e = 1$. In these units the velocity of light in vacuum $c$ numerically coincides with 
the inverse value of the dimensionless fine structure constant, i.e. $c = \alpha^{-1}$, where $\alpha = \frac{e^2}{\hbar c} \approx$ 7.2973525698$\cdot 10^{-3} 
\approx$ 1/137.035999074 is the fine structure constant \cite{CRC}. In atomic units the same Hamiltonian, Eq.(\ref{Ham}), is written in the form
\begin{equation}
 H = -\frac12 \Bigl( \nabla^{2}_{1} + \nabla^{2}_{2} + \frac{1}{M_N} \nabla^{2}_{3} \Bigr) - \frac{Q}{r_{32}} - \frac{Q}{r_{31}} + \frac{1}{r_{21}} \label{Ham1}
\end{equation}

It should be emphasized that our approach based on the use of non-relativistic wave functions will work, if (and only if) the non-relativistic variational wave 
functions have been determined to very high accuracy (precise wave functions). For the ground $1^{1}S(L = 0)-$states of the light two-electron ions the highly accurate 
wave functions are approximated with the use of the exponential variational expansion in relative coordinates $r_{32}, r_{31}$ and $r_{21}$ (see, e.g., \cite{Fro98} 
and references therein)
\begin{eqnarray}
 \Psi = \Bigl( 1 + \hat{P}_{12} \Bigr) \sum_{i=1}^{N}  C_{i} \exp(-\alpha_{i} r_{31} - \beta_{i} r_{31} - \gamma_{i} r_{21})  \label{exp1} 
\end{eqnarray}
Each of these three relative coordinates $r_{ij}$ is defined as the difference between the corresponding Cartesian coordinates of the two particles, e.g., $r_{ij} = 
\mid {\bf r}_i - {\bf r}_j \mid$, where ${\bf r}_i$ and ${\bf r}_j$ are the Cartesian coordinates of particles $i$ and $j$. It follows from this definition that the 
relative coordinates $r_{32}, r_{31}$ and $r_{21}$ are translationally and rotationally invariant. The coefficients $C_i$ are the linear (or variational) parameters 
of the variational expansion, Eq.(\ref{exp1}), while the parameters $\alpha_{i}, \beta_{i}$ and $\gamma_{i}$ are the non-linear (or varied) parameters of this 
expansion. To optimize such non-linear parameters in Eq.(\ref{exp1}) we have developed a very effective two-stage optimiztion strategy \cite{Fro98}. The operator 
$\hat{P}_{12}$ in Eq.(\ref{exp1}) is the permutation operator for two identical particles (electrons). 

In this study we consider several light two-electron (or He-like) ions: Li$^{+}$, Be$^{2+}$, B$^{3+}$ and C$^{4+}$. Our results given in Tables I - IV 
allow one to determine the normal and specific components of the isotopic shifts $\Delta E_M$ in these cases. Furthermore, by using the expectation value 
of the electron-nucleus delta-functions in each of these ions one can determine the corresponding field shifts $\Delta E_{F}$ (see Table V). Formally, our 
data from Tables I - V contain all expectation values which are needed to determine the numerical values of $\Delta E_M$ and $\Delta E_F$. All data 
presented in these Tables correspond to the two-electron ions with infinitely heavy nuclei. The significance of the computed components of the isotopic shift 
($\Delta E_M$ and $\Delta E_F$) in actual applications can be reliably determined, if we can evaluate the same expectation values for the atoms/ions with 
the finite nuclear masses. This problem is discussed in Section IV. 

Here we want to consider another problem which arises during numerical evaluation of the $\Delta E_M$ and $\Delta E_F$ components by using our formulas. Indeed, 
as we mentioned above, numerical evaluations of the isotope shifts in these two-electron ions is reduced to highly accurate calculations of the expectation values 
of some electron-nucleus and electron-electron operators, which include the inter-particle delta-functions. The overall accuracy of these expectation values is a 
crucial question for highly accurate evaluations of the isotope shifts. In turn, this is directly related to the overall accuracy of the wave functions used in 
calculations. Since the early years of quantum mechanics the accuracy of the variational wave functions has been assessed by minimizing the total energy computed 
with such a wave function. This simple `criterion of the quality' does not work for the expectation values of operators which are needed to determine the isotopic 
shifts in light ions/atoms. Indeed, currently by using a number of special methods, e.g., GFMC method (or Green Function Monte-Carlo method), it is easy to 
construct bound state wave functions which produce `essentially exact' total energies for different few-body systems, but the expectation values of some other 
properties computed with such wave functions are relatively inaccurate. In particular, it is difficult to determine highly accurate expectation values of the 
electron-nucleus and electron-electron delta-functions, i.e. the local properties, or properties determined at one spatial point. Therefore, below we need to 
discuss numerical criteria that are used to judge the overall quality of the wave functions and allow us to evaluate the applicability of these wave functions for 
accurate numerical computations of all interparticle delta-functions and other local properties some of which also include spatial derivatives of different orders. 
The second closely related question is the convergence rate (upon the total number $N$ of basis functions used) for the expectation values which are needed for 
numerical evaluation of the isotope shifts. These two questions are considered in this Section. 
 
A natural criterion of the quality of the wave functions which was used already in the first variational calculations of atomic and molecular systems is based 
on the virial theorem (see, e.g., \cite{Fock}). The virial theorem for Coulomb systems, e.g., for atoms and ions, can be written in the form $2 \langle T \rangle 
= - \langle V \rangle$, where $T$ is the operator of the kinetic energy and $V$ is the operator of the potential energy. Since the Hamiltonian $H$ is 
represented as the sum $H = T + V$, then one finds for the expectation values $\langle H \rangle = E = - \langle T \rangle = \frac12 \langle V \rangle$, where 
$E$ is the total energy of the atomic system bound by the Coulomb interparticle potentials. In general, this criterion is simple, but in many cases is not 
sufficient to evaluate the overall quality of the variational wave functions which then can be used for various purposes, e.g., to determine the expectation 
values of different quantum operators. It is clear that some other criteria are needed. Fortunately, for all Coulomb few-body systems we can always evaluate 
(exactly) the particle densities at the two-particle coalescence points. For instance, for the three-particle (or two-electron) ions Li$^{+}$, Be$^{2+}$, B$^{3+}$ 
and C$^{4+}$ we have two such a coalescence points: the electron-nucleus point and electron-electron point. At each of these points we can calculate the 
expectation values of the following operators (or cusp-operators)
\begin{eqnarray}
 \nu_{eN} = \frac{\langle \delta({\bf r}_{eN}) \frac{\partial}{\partial r_{eN}} \rangle}{\langle \delta({\bf r}_{eN}) \rangle} 
 \label{eqf6}
\end{eqnarray}
in the case of the electron-nucleus cusp, and
\begin{eqnarray}
 \nu_{ee} = \frac{\langle \delta({\bf r}_{ee}) \frac{\partial}{\partial r_{ee}} \rangle}{\langle \delta({\bf r}_{ee}) \rangle} 
 \label{eqf61}
\end{eqnarray}
for the electron-electron cusp. These two expectation values must coincide with the known values of these cusps, i.e., with the following numerical values (in atomic 
units)
\begin{eqnarray}
 \nu_{eN} = - Q e^2 \frac{m_e M_N}{m_e + M_N} = - Q \frac{1}{1 + \frac{1}{M_N}} = -Q \Bigr( 1 + M^{-1}_N \Bigl)^{-1} \; \; \; , \; \; \; \nu_{ee} = 0.5 
 \; \; \; \label{eqf62}
\end{eqnarray}
where $M_N = \frac{M_N}{m_e}$ is the nuclear mass which can be finite (real), or infinite for model atomic systems.  

The coincidence of these two expectation values, Eqs.(\ref{eqf6}) - (\ref{eqf61}), with the predicted values, Eq.(\ref{eqf62}), is a very effective test for the
variational wave functions in any Coulomb system. In real applications, however, different authors try to `improve' their actual cusp values using various tricks, 
e.g., by adding additional `special' terms to the wave functions. These additional terms do not change the computed variational energy, but they allow one to obtain 
`very accurate' cusps. Such results are published by some authors to support claims of extremely high quality of variational wave functions. In general, 
at this moment it is hard to trust such results without investigating expectation values of other quantum operators. On the other hand, for each of the 
two-electron atoms and ions (at least, for their ground and low-excited states) one finds in the literature a large number of expectation values already 
computed to high numerical accuracy. Formally, all conclusions about the overall quality of the three-particle wave function must follow from numerical 
comparison of the computed expectation values and values known from the literature. Ideally, such a complete set of required expectation values includes not 
only regular expectation values, but singular expectation values too. For actual two-electron ions (and, in general, for few-electron atomic systems) it is 
possible to find a few such properties (or `complex' expectation values) which include the expectation values of all delta-functions and at least one singular 
expectation value. Accurate numerical computations of this set of expectation values provide a very effective criterion for the overall quality of the wave 
function used. 

In reality, a number of such criteria for the quality of the variational wave functions used in calculations can be found among various lowest-order relativistic 
and QED corrections. Indeed, these corrections contain singular expectation values and different interparticle delta-functions. By determining these corrections and 
comparing results with the known numerical values one can estimate the quality of the trial wave functions. In particular, below we determine the lowest-order QED 
correction for each of the two-electron ions considered. However, at the first stage of our calculations we apply the electron-nuclear and electron-electron cusps 
as criteria of the quality of our wave functions.

In this study we determine both the electron-nucleus and electron-electron cusp values and compare them with the known (or expected) cusps, i.e., with $-Q 
\Bigr( 1 + M^{-1}_N \Bigl)^{-1}$ and 0.5 (in atomic units), respectively. In calculations performed here we have also determined many dozens of other
expectation values, including some singular expectation values (see the corresponding Tables in \cite{Fro2006}). In general, all these expectation values are 
very close to the values given in \cite{Fro2006}, but our current values are more accurate. This means that our wave functions have better numerical accuracy. 
Numerical values of the paticular expectation values (in atomic units) needed for numerical evaluations of the isotopic shifts in the Li$^{+}$, Be$^{2+}$, B$^{3+}$ 
and C$^{4+}$ ions with the infinitely heavy nuclei are presented in Tables I - IV, respectively. Expectation values from Tables  I - IV include the total energies, 
electron-nucleus delta-functions and cusp values. Each of these Tables also contain the expectation values $\frac12 \langle {\bf p}^2_1 \rangle, \langle {\bf p}_1 
\cdot {\bf p}_2 \rangle$ and $\frac12 \langle {\bf p}^2_N \rangle$ which are needed for zero-order evaluation of the isotopic shifts. Based on these results and by 
applying the formula, Eq.(\ref{eqf5}), we have determined the numerical values of the field components of isotopic shifts which are presented in Table V (in atomic 
units). This Table also includes numerical values of the following factors from the formula, Eq.(\ref{eqf5}): $R$ (the actual nuclear radius), $b, X = 4^{b+1} \pi 
Q^{2b-1} \cdot \alpha^{4 b} \cdot \frac{3 (b + 1)}{[\Gamma(2 b + 1)]^2 (2 b + 1) (2 b + 3)}$ and $Y = \Bigl( \frac{R}{r_e} \Bigr)^{2 b}$. The expectation values of 
the electron-nuclear delta-functions were taken from Tables I - IV. To evaluate the Euler's gamma-function $\Gamma(x)$ we have used approximate 7-term formula 
derived by Lanczos \cite{Lanc}. Finally, the overall accuracy of our formula for $\Delta E_F$ has been estimated as $\approx 1 \cdot 10^{-10} - 2 \cdot 10^{-10}$ 
$a.u.$ 

\section{Two-electron ions with the finite-mass nuclei}

Tables I - V contain results of numerical calculations in which the masses of all atomic nuclei were assumed to be infinite. For light atoms and ions numerical 
errors related with the finite nuclear masses can be substantial. Even in those cases, when all finite-mass corrections have been evaluated and included in the 
final formulas one can still identify some numerical errors in the total energies and other properties since such errors can easily be detected in modern highly 
accurate calculations. Moreover, it is hard to compare directly our computational data obatained for the model ions with infinitely heavy nuclei with the results of 
precise optical observations performed for actual ions with the finite nuclear masses. The overall accuracy of modern optical experiments based on the use of lasers 
is already extremely high and continues to increase. Biefly, this means that numerical calculations based on the use of small parameter(s) and perturbation theory 
lead to a very complex procedure which cannot provide a very high accuracy for the final results. An alternative way is to perform all calculations for the two-electron 
ions which have the finite nuclear masses from the first step of the procedure. In other words, we need to consider a general three-body problem for Coulomb systems. In 
this approach we re-calculated the results from Tables I - V for a number of actual ions, i.e. for the Li$^{+}$, Be$^{2+}$, B$^{3+}$ and C$^{4+}$ ions with the finite 
nuclear masses. The nuclear masses of the different Li, Be, B and C isotopes used in these calculations are (in atomic units): $M({}^{6}$Li) = 10961.8968 $m_e$, 
$M({}^{7}$Li) = 12786.3927 $m_e$, $M({}^{9}$Be) = 16424.2032 $m_e$, $M({}^{10}$Be) = 18249.5555 $m_e$, $M({}^{10}$B) = 18247.4677 $m_e$, $M({}^{11}$B) = 20063.7631 
$m_e$, $M({}^{12}$C) = 21868.66182 $m_e$, $M({}^{13}$C) = 23697.66580 $m_e$ and $M({}^{14}$C) = 25520.34677 $m_e$.The total energies of these ions with the finite 
nuclear masses can be found in Table VI.

The expectation values of the operators which are needed to determine the isotopic shifts (or any mass-related shifts) in these two-electron ions can be found in 
Table VII. These operators are: ${\bf p}^2_1, {\bf p}_1 \cdot {\bf p}_2, {\bf p}^2_N$ and $\delta({\bf r}_{eN})$. For instance, let us assume that in our calculations 
of some of these two-electron ions we have used the nuclear mass $M$, while in new experiemts it was found that such a mass equals $M^{\prime}$. The corresponding 
correction (or additional mass correction $\Delta^{(A)}_{M}$) is written in the form
\begin{eqnarray}
 \Delta^{(A)}_{M} = \Bigl( \frac{1}{2 M^{\prime}} - \frac{1}{2 M} \Bigr) \langle {\bf p}^2_N \rangle = \Bigl( \frac{1}{M^{\prime}} - \frac{1}{M} \Bigr) 
 (\langle {\bf p}^2_1 \rangle + \langle {\bf p}_1 \cdot {\bf p}_2 \rangle) \; \; \; \label{equ1}
\end{eqnarray}
where $\langle {\bf p}^2_N \rangle, \langle {\bf p}^2_1 \rangle$ and $\langle {\bf p}_1 \cdot {\bf p}_2 \rangle$ expectation values must be determined for the atomic
system with the finite nuclear mass $M$ (i.e. `old' nuclear mass). Analogously, by using the expectation values of the electron-nucleus delta-functions from Table 
VII one can evaluate the corresponding `additional' field shift $\Delta E^{(A)}_{F}$ which is related to the `new' nuclear radius $R^{\prime}_N$ measured in the 
experiments
\begin{eqnarray}
 \Delta E^{(A)}_{F} = \frac{4 \pi a^2_0}{Q} \cdot \frac{b + 1}{[\Gamma(2 b + 1)]^2} \cdot B(b) \cdot \Bigl( \frac{2 Q}{a_0} \Bigr)^{2 b} \cdot 
  (R^{\prime} - R) \cdot \langle \delta({\bf r}_{eN}) \rangle \label{equ2}
\end{eqnarray} 
where $R$ designates the `old' nuclear radius. In reality, all nuclear radii are currently known to a numerical accuracy $\approx 1 \cdot 10^{-3}$ $fm$ (and even 
better). Therefore, the absolute values of the differences $R^{\prime} - R$ are very small ($R^{\prime} - R \le 1 \cdot 10^{-3} R$ (and smaller)). 

The expectation values of operators from Table VII allow one to determine and evaluate the actual and `additional' mass and field shifts for each of the two-electron 
ions considered in this study. However, for each atomic system one also finds a separate group of small corrections which must be added to the computed total energy, 
or isotopic shift. One group of such small corrections  ($\simeq \alpha^{2}$) is directly related to the fact that $\frac{v_e}{c}$ is not zero exactly (it is small 
$\approx 1 \cdot 10^{-5}$, but non-zero!). Another group of small corrections ($\simeq \alpha^{3}$) to the total energies arises from interaction between atomic 
electron(s) and radiation quanta. These group of corrections is called the lowest-order QED corrections. These corrections can be determined to high numerical 
accuracy with the use of a few expectation values computed with the non-relativistic wave functions. Both these corrections are discussed below.     

\section{Relativistic corrections}

The results of highly accurate calculations for two-electron atoms and ions are of great interest by themselves. However, in the Sections above we considered only the 
non-relativistic total energies of a few two-electron ions in their ground $1^1S-$states. These energies were determined to high numerical accuracy for model 
ions with the infinitely heavy nuclei and also for actual ions with the finite nuclear masses $M_N$. It appears that the total energies and all expectation values computed 
with the non-relativistic wave functions are analytical functions of the dimensionless parameter $\frac{m_e}{M_N} = \frac{1}{M_N}$ which is the inverse mass of the 
nucleus expressed in the electron mass $m_e$. In reality, such an idealized picture ignores all lowest-order relativistic and quantum electrodynamic corrections (or QED 
corrections, for short) for actual two-electron ions. For light two-electron ions these corrections are relatively small $\approx 10^{-3}$ - $10^{-6}$ $a.u.$ (the 
overall values are different for different states and ions), but they can be important in some problems. Here we want to restrict our analysis to the lowest-order 
relativistic ($\simeq \alpha^2$) and QED ($\simeq \alpha^3$) corrections to the energy levels. In this Section we discuss numerical calculations of the lowest-order 
relativistic corrections. The lowest-order QED corrections are considered in the next Section.

The general theory developed for numerical evaluation of the relativistic corrections in light two-electron atoms and ions can be found in \cite{BS} and for 
many-electron atoms in \cite{FrFi}. For the ground (singlet) $1^1S-$states in these ions one finds a number of significant simplifications in the general theory. To 
simplify our analysis even further here we write only the final formula which is used to determine the lowest-order relativistic correction in the ground singlet 
state of two-electron atoms/ions. In atomic units this formula takes the form 
\begin{eqnarray}
 \Delta E_{R} = - \frac{\alpha^2}{4} \langle {\bf p}^4_1 \rangle - \frac{\alpha^2}{2} \langle \frac{1}{r_{12}} {\bf p}_1 \cdot {\bf p}_2 \rangle - 
  \frac{\alpha^2}{2} \langle \frac{1}{r^{3}_{12}} {\bf r}_{12} ({\bf r}_{12} \cdot {\bf p}_1) {\bf p}_2 \rangle 
 + \pi \alpha^2 \Bigl[ Q \langle \delta({\bf r}_{eN}) \rangle + \langle \delta({\bf r}_{ee}) \rangle \Bigr] \; \; \; \;  \label{relat}
\end{eqnarray}
where $\alpha = 7.2973525698 \cdot 10^{-3}$ is the fine-structure constant and $Q$ is the electric charge of the nucleus expressed in terms of the electron charge $e$. 
In this equation the particles with indexes 1 and 2 are the electrons, while the particle with index 3 is the atomic nucleus. The notation $r_{ee} = r_{12}$ stands for the 
electron-electron distance, while the notation $r_{eN} = r_{13} (= r_{23})$ designates the electron-nuclear distance. The formula, Eq.(\ref{relat}), can be used to 
determine the lowest-order ($\sim \alpha^2$) relativistic correction to the non-relativistic total energies of the ground $1^1S-$states in light two-electron ions. For 
the two-electron ions considered in this study, the last term (i.e. the sum of the expectation values of the delta-functions) in Eq.(\ref{relat}) equals: 
3.528190130474$\cdot 10^{-3}$ $a.u.$ for the ${}^{\infty}$Li$^{+}$ ion, 1.176338007844$\cdot 10^{-2}$ $a.u.$ for the ${}^{\infty}$Be$^{2+}$ ion,  
2.962880936734$\cdot 10^{-2}$ $a.u.$ for the ${}^{\infty}$B$^{3+}$ ion and 6.270223554881$\cdot 10^{-2}$ $a.u.$ for the ${}^{\infty}$C$^{4+}$ ion. The expectation values 
of other terms from Eq.(\ref{relat}) will be presented elsewhere. 

\section{The lowest order QED correction}

As we mentioned above actual calculations of all relativistic and QED corrections in two-electron atoms and ions can be performed only with the use of the truly 
relativistic wave functions. The non-relativistic wave functions obtained as the solutions of the Schr\"{o}dinger equation can be applied only for numerical evaluations 
of some lowest-order relativistic and QED corrections in light two-electron ions. Below, we discuss numerical evaluation of the lowest order QED correction $\Delta E^{QED}$ 
for the two-electron ions: Li$^{+}$, Be$^{2+}$, B$^{3+}$ and C$^{4+}$. The corresponding formula for such a correction $\Delta E^{QED}$ in a two-electron ion with infinitely 
heavy nucleus is written in the form (in atomic units) 
\begin{eqnarray}
 \Delta E^{QED} &=& \frac{8}{3} Q \alpha^3 \Bigl[ \frac{19}{30} - 2 \ln \alpha - \ln K_0 \Bigr]
 \langle \delta({\bf r}_{eN}) \rangle 
 + \alpha^3 \Bigl[ \frac{164}{15} + \frac{14}{3} \ln \alpha - \frac{10}{3} S(S + 1) \Bigr]
 \langle \delta({\bf r}_{ee}) \rangle \nonumber \\
 &-& \frac{14}{3} \alpha^3 \langle \frac{1}{r^{3}_{ee}} \rangle \label{eqf8}
\end{eqnarray}
where $\alpha$ is the fine structure constant, $Q$ is the nuclear charge (in atomic units) and $S$ is the total electron spin. The ground states in all two-electron ions 
considered in this study are the singlet states with $S = 0$. Also, in this formula $\ln K_0$ is the Bethe logarithm. To determine the Bethe logarithm which is usually
evaluated by applyin the formula $\ln K_0 = \ln k_0 + 2 \ln Q$, where $\ln k_0$ is the charged-reduced Bethe logarithm. Numerical values of the Bethe logarithm were 
evaluated for each of these two-electron ions (Li$^{+}$, Be$^{2+}$, B$^{3+}$ and C$^{4+}$) in earlier works. 

The last term in Eq.(\ref{eqf8}) is called the Araki-Sucher term, or Araki-Sucher correction, since this correction was obtained and investigated for the first time in 
papers by Araki and Sucher \cite{Araki}, \cite{Such}. The expectation value of the term $\langle \frac{1}{r^{3}_{ee}} \rangle$ is singular, i.e., it contains the regular 
and non-zero divergent parts. A general theory of singular exponential integrals was developed in our earlier works (see, e.g., \cite{Fro2005} and references therein). 
In particular, in \cite{Fro2005} we have shown that the $\langle \frac{1}{r^{3}_{ee}} \rangle$ expectation value is determined by the formula
\begin{eqnarray}
  \langle \frac{1}{r^{3}_{ee}} \rangle = \langle \frac{1}{r^{3}_{ee}} \rangle_R + 
  4 \pi \langle \delta({\bf r}_{ee}) \rangle
\end{eqnarray} 
where $\langle \frac{1}{r^{3}_{ee}} \rangle_R$ is the regular part of this expectation value and $\langle \delta({\bf r}_{ee}) \rangle$ is the expectation value of the 
electron-electron delta-function. Briefly, we can say that the overall contribution of the singular part of the $\frac{1}{r^{3}_{ee}}$ operator is reduced to the expectation 
value of the corresponding delta-function. An analogous formula can be written for the $\langle \frac{1}{r^{3}_{eN}} \rangle$ expectation value. By using the data from Table 
VII we determine the lowest order QED corrections $\Delta E^{QED}$ for the ground $1^1S-$states in the ${}^{\infty}$Li$^{+}$, ${}^{\infty}$Be$^{2+}$, ${}^{\infty}$B$^{3+}$ 
and ${}^{\infty}$C$^{4+}$ two-electron ions. These values are: $\Delta E^{QED}($Li$^{+})$ = 1.102 475 518$\cdot 10^{-4}$ $a.u.$ (or 7.253 938 191$\cdot 10^{5}$ $MHz$),   
$\Delta E^{QED}($Be$^{2+})$ = 3.303 586 656$\cdot 10^{-4}$ $a.u.$ (or 21.736 555 344$\cdot 10^{5}$ $MHz$),
$\Delta E^{QED}($B$^{3+})$ = 7.581 229 698$\cdot 10^{-4}$ $a.u.$ (or 49.882 095 145$\cdot 10^{5}$ $MHz$) and 
$\Delta E^{QED}($C$^{4+})$ = 14.744 207 461$\cdot 10^{-4}$ $a.u.$ (or 97.012 224 754$\cdot 10^{5}$ $MHz$). To re-calculate the data from atomic units to $MHz$ we used the most 
recent conversion factor from $a.u.$ to $MHz$ which equals 6.579 683 920 729$\cdot 10^{9}$.

For two-electron ions with finite nuclear mass we need to evaluate the corresponding recoil correction to the lowest-order QED correction. Such a correction is also given 
in \cite{Fro2006}. In atomic units it is written in the following form
\begin{eqnarray}
 \Delta E^{QED}_{M} &=& \Delta E^{QED}_{\infty} - \Bigl(\frac{2}{M} + \frac{1}{M + 1} \Bigr)
 \Delta E^{QED}_{\infty} + \frac{4 \alpha^3 Q^2}{3 M} \Bigl[ \frac{37}{3} - \ln \alpha - 4 \ln K_0 \Bigr]
 \langle \delta({\bf r}_{eN}) \rangle \nonumber \\
 &+& \frac{7 \alpha^3}{3 \pi M} \langle \frac{1}{r^{3}_{eN}} \rangle \label{eqf9}
\end{eqnarray} 
where $M \gg m_e$ is the nuclear mass. All expectation values in this equation must be determined for the real two-electron ions which have the finite nuclear masses. The inverse 
mass $\frac{1}{M}$ is a small dimensionless parameter which for the ions considered in this study  is $\le 1 \cdot 10^{-4}$. By using our expectation values for 
the electron-nucleus and electron-electron delta-functions and for the corresponding Araki-Sucher terms ($\langle \frac{1}{r^{3}_{eN}} \rangle$ and $\langle \frac{1}{r^{3}_{ee}} 
\rangle$) we have determined the lowest order QED corrections for each of the ions considered in this study. Numerical values of these lowest-order QED corrections (in atomic 
units) can be found in Table VIII. 

\section{Conclusion}

We have performed highly accurate computations of the ground $1^1S-$states in four two-electron, light ions: Li$^{+}$, Be$^{2+}$, B$^{3+}$ and C$^{4+}$. The results of our
calculations allow us to evaluate (to very high accuracy) the non-relativistic isotopic shifts for a number of isotopes of these ions. The expectation values of different 
operators, which are needed during this procedure, have been determined to very high accuracy and allow one to evaluate the non-relativistic isotopic shifts for all isotopes of 
the two-electron, light ions discussed in this study. The lowest-order QED corrections (Quantun Elelctrodynamics corrections) have been also evaluated to high numerical accuracy 
for each isotope of the four two-electron ions. We also discuss formulas which will be used in our next study to perform numerical calculations of the lowest-order relativistic 
corrections for the ground states of these two-electron ions. Future plans also include improvement of our old method which was used earlier for numerical calculations of Bethe 
logarithm.  

\begin{center}
 {\bf Appendix. Weiz\"{a}cker mass formula}
\end{center}

The formula which provides a uniform relation between the nuclear mass $M$ and total number of nucleons $A$, nuclear charge $Z$ (= number of protons $N_p$) and number of neutrons 
$N_n$ in the nucleus was derived in 1937 by Bethe, Weiz\"{a}cker and others (known as the Weiz\"{a}cker formula \cite{Weiz}, or Bethe-Weiz\"{a}cker formula). This five-term formula 
for the nuclear binding energy $E_b$ was produced 75 years ago and since then its general structure has never been changed. First, note that the mass formula for an arbitrary nucleus 
can be written in the form 
\begin{eqnarray}
 M = m_p \Bigl[ Z + N \Bigl( \frac{m_n}{m_p} \Bigr) - \frac{E_b}{m_p c^2} \Bigr] 
 \label{eap1}
\end{eqnarray}
where $M$ is the nuclear mass of the nucleus with $A$ nucleons, $Z$ protons and $N$ neutrons, i.e. $A = Z + N$. Also in this formula $E_b$ is the binding energy of the nucleus,
$c$ is the speed of light in vacuum, while $m_p$ and $m_n$ are the masses of the proton and neutron, respectively. Based on the results of most recent experiments we have 
for the factors $m_p c^2$ = 938.272910 $MeV$ and $m_n c^2$ = 939.565378 $MeV$. The advantage of the formula, Eq.(\ref{eap1}), is obvious, since it contains only dimensionless ratios 
and two integer numbers ($Z$ and $N$). For instance, if we choose in Eq.(\ref{eap1}) $m_p = 1836.152701 m_e$, then $M$ will be given in $m_e$ (or in atomic units if $m_e = 1$). This 
is very convenient for highly accurate computations of different few-electron ions. 

The parameter $E_b$ in Eq.(\ref{eap1}) is called the binding energy of the nucleus. The explicit expression for the nuclear binding energy $E_b$ is written as the following 
sum (the Weiz\"{a}cker formula):
\begin{eqnarray}
 E_b = a_V A - a_S A^{\frac23} - a_C \frac{Z^2}{A^{\frac13}} - a_A \frac{(N - Z)^2}{A} + \delta(A,Z) \label{eap2}
\end{eqnarray}
where the five terms in the right-hand side of this equation are called the volume term, surface term, Coulomb term, asymmetry term and pairing term, respectively. The 
pairing term $\delta(A,Z)$ equals zero, if $A$ is odd. If $A$ is even and both $Z$ and $N$ are even, then $\delta(A,Z) = \frac{a_p}{\sqrt{A}}$. The Weiz\"{a}cker formula 
is relatively accurate for regular nuclei, i.e. for nuclei which are not far from the center of the stability region. In reality, such an accuracy directly depends upon 
the numerical values of parameters $a_V, a_S, a_C, a_A$ and $a_p$ in Eq.(\ref{eap2}). To obtain the lowest-order approximation in our calculations we have used the following 
values of these parameters: $a_V$ = 15.8 $MeV$, $a_S$ = 18.3 $MeV$, $a_C$ = 0.714 $MeV$, $a_A$ = 23.2 $MeV$ and $a_p$ = 12.0 $MeV$. For all heavy nuclei and even for carbon 
nuclei the overall accuracy of Weiz\"{a}cker mass formula is sufficient to determine the mass of the nucleus which can later be used to perform highly accurate atomic 
calculations. However, for Li-atoms, Be-like and B-like ions the numerical values of these parameters in the Weiz\"{a}cker mass formula \cite{Weiz} must be modified. The reason 
for this follows from the fact that Weiz\"{a}cker mass formula ignores the actual shell structure which is of great importance for light atomic nuclei and, therefore, it is not 
accurate for some light nuclei, e.g., for all nuclei of the hydrogen and helium isotopes.

\newpage
\begin{table}[tbp]
   \caption{The total energies $E$ and expectation values of the electron-nuclear delta-function
            $\delta_{eN}$, electron-nuclear cusp $\nu_{eN}$ and some other operators for the 
            two-electron lithium ion Li$^{+}$ (in atomic units). $K$ is the total number of 
            basis functions used.}
     \begin{center}
     \begin{tabular}{| c | c | c | c |}
      \hline\hline
 $K$ & $E$(Li$^{+}$) & $\langle \delta_{eN} \rangle$ & $\nu_{eN}$ \\
     \hline
 3500 & -7.279913 412669 305964 918264 &  6.8520 094343 431 &  -3.0000 00000 158 \\
 
 3700 & -7.279913 412669 305964 918525 &  6.8520 094343 456 &  -3.0000 00000 125 \\

 3840 & -7.279913 412669 305964 918626 &  6.8520 094343 460 &  -2.9999 99999 918 \\

 4000 & -7.279913 412669 305964 918727 &  6.8520 094343 462 &  -2.9999 99999 901 \\
     \hline
 $K$ & $\frac12 \langle {\bf p}^2_1 \rangle$ &  $\langle {\bf p}_1 \cdot {\bf p}_2 \rangle$ &  $\frac12 \langle {\bf p}^2_N \rangle$ \\
      \hline\hline
 3500 & 3.63995 670633 465298 240 &  0.288975 786393 989535 661 &  7.56888 919906 329532 141 \\
 
 3700 & 3.63995 670633 465298 241 &  0.288975 786393 989535 661 &  7.56888 919906 329532 143 \\

 3840 & 3.63995 670633 465298 241 &  0.288975 786393 989535 662 &  7.56888 919906 329532 144 \\

 4000 & 3.63995 670633 465298 242 &  0.288975 786393 989535 662 &  7.56888 919906 329532 145 \\
    \hline\hline
  \end{tabular}
  \end{center}
  \end{table}
\begin{table}[tbp]
   \caption{The total energies $E$ and expectation values of the electron-nuclear delta-function
            $\delta_{eN}$, electron-nuclear cusp $\nu_{eN}$ and some other operators for the 
            two-electron berillium ion Be$^{2+}$ (in atomic units). $K$ is the total number of 
            basis functions used.}
     \begin{center}
     \begin{tabular}{| c | c | c | c |}
      \hline\hline
 $K$ & $E$(Be$^{2+}$) & $\langle \delta_{eN} \rangle$ & $\nu_{eN}$ \\
     \hline
 3500 & -13.65556 623842 358670 207905 &  17.1981 72544 645 &  -3.9999 99999 962 \\
 
 3700 & -13.65556 623842 358670 207949 &  17.1981 72544 640 &  -3.9999 99999 921 \\

 3840 & -13.65556 623842 358670 207968 &  17.1981 72544 638 &  -4.0000 00000 125 \\

 4000 & -13.65556 623842 358670 207994 &  17.1981 72544 635 &  -4.0000 00000 148 \\
     \hline
 $K$ & $\frac12 \langle {\bf p}^2_1 \rangle$ &  $\langle {\bf p}_1 \cdot {\bf p}_2 \rangle$ &  $\frac12 \langle {\bf p}^2_N \rangle$ \\
      \hline\hline
 3500 & 6.82778 311921 179335 084 &  0.420520 303439 441862 011 &  14.07608 654186 302856 368 \\ 
 
 3700 & 6.82778 311921 179335 086 &  0.420520 303439 441862 010 &  14.07608 654186 302856 369 \\

 3840 & 6.82778 311921 179335 089 &  0.420520 303439 441862 009 &  14.07608 654186 302856 370 \\

 4000 & 6.82778 311921 179335 091 &  0.420520 303439 441862 009 &  14.07608 654186 302856 371 \\
    \hline\hline
  \end{tabular}
  \end{center}
  \end{table}
 \begin{table}[tbp]
   \caption{The total energies $E$ and expectation values of the electron-nuclear delta-function
            $\delta_{eN}$, electron-nuclear cusp $\nu_{eN}$ and some other operators for the 
            two-electron boron ion B$^{3+}$ (in atomic units). $K$ is the total number of basis 
            functions used.}
     \begin{center}
     \begin{tabular}{| c | c | c | c |}
      \hline\hline
  $K$ & $E$(B$^{3+}$) & $\langle \delta_{eN} \rangle$ & $\nu_{eN}$ \\
     \hline\hline
 3500 & -22.03097 1580242 781541 654073 &  34.758 743660 955 &  -5.0000 0000 319 \\
 
 3700 & -22.03097 1580242 781541 654321 &  34.758 743660 965 &  -5.0000 0000 235 \\

 3840 & -22.03097 1580242 781541 654548 &  34.758 743660 947 &  -5.0000 0000 107 \\

 4000 & -22.03097 1580242 781541 654663 &  34.758 743660 935 &  -5.0000 0000 119 \\
      \hline
 $K$ & $\frac12 \langle {\bf p}^2_1 \rangle$ &  $\langle {\bf p}_1 \cdot {\bf p}_2 \rangle$ &  $\frac12 \langle {\bf p}^2_N \rangle$ \\
      \hline\hline
 3500 & 11.01548 579012 139077 100 &  0.552752 631642 101467 789 &  22.58372 421188 488300 979 \\ 
 
 3700 & 11.01548 579012 139077 089 &  0.552752 631642 101467 734 &  22.58372 421188 488300 952 \\

 3840 & 11.01548 579012 139077 086 &  0.552752 631642 101467 715 &  22.58372 421188 488300 942 \\

 4000 & 11.01548 579012 139077 083 &  0.552752 631642 101467 701 &  22.58372 421188 488300 938 \\
     \hline\hline
  \end{tabular}
  \end{center}
  \end{table}
  \begin{table}[tbp]
   \caption{The total energies $E$ and expectation values of the electron-nuclear delta-function
            $\delta_{eN}$, electron-nuclear cusp $\nu_{eN}$ and some other operators for the 
            two-electron carbon ion C$^{4+}$ (in atomic units). $K$ is the total number of basis 
            functions used.}
     \begin{center}
     \begin{tabular}{| c | c | c | c |}
      \hline\hline
 $K$ & $E$(C$^{4+}$) & $\langle \delta_{eN} \rangle$ & $\nu_{eN}$ \\
     \hline\hline
 3500 & -32.40624 660189 853031 055622 &  61.443 578056 445 &  -5.9999 99998 765 \\
 
 3700 & -32.40624 660189 853031 055638 &  61.443 578056 514 &  -5.9999 99999 871 \\

 3840 & -32.40624 660189 853031 055647 &  61.443 578056 537 &  -6.0000 00000 048 \\ 

 4000 & -32.40624 660189 853031 055660 &  61.443 578056 543 &  -6.0000 00000 037 \\ 
      \hline
 $K$ & $\frac12 \langle {\bf p}^2_1 \rangle$ & $\langle {\bf p}_1 \cdot {\bf p}_2 \rangle$ & $\frac12 \langle {\bf p}^2_N \rangle$ \\
      \hline\hline
 3500 & 16.20312 330094 926515 523 &  0.685334 822135 598924 527 &  33.09158 142403 412923 500 \\ 
 
 3700 & 16.20312 330094 926515 524 &  0.685334 822135 598924 535 &  33.09158 142403 412923 502 \\

 3840 & 16.20312 330094 926515 525 &  0.685334 822135 598924 536 &  33.09158 142403 412923 502 \\

 4000 & 16.20312 330094 926515 525 &  0.685334 822135 598924 537 &  33.09158 142403 412923 503 \\
     \hline\hline
  \end{tabular}
  \end{center}
  \end{table}
  \begin{table}[tbp]
   \caption{The nuclear radius $R$ ($fm$), parameter $b$, factors $X$ and $Y$ (see the main 
            text) and field components of the total isotopic shift $\Delta E_{F}$ (all values 
            are in atomic units) for each isotope.}
     \begin{center}
     \begin{tabular}{| c | c | c | c | c | c | c |}
      \hline\hline
 isotope & $Q$ & $R$ & $b$ & X & Y & $\Delta E_{F}$ \\
     \hline\hline 
 ${}^{6}$Li  & 3 & 2.5385 & 0.99976034018621 & 4.297056149289$\cdot 10^{-8}$ & 0.81154478 & 2.388077748$\cdot 10^{-7}$ \\

 ${}^{7}$Li  & 3 & 2.4312 & 0.99976034018621 & 4.297056149289$\cdot 10^{-8}$ & 0.74440369 & 2.191262432$\cdot 10^{-7}$ \\
       \hline
 ${}^{9}$Be  & 4 & 2.5190 & 0.99957389838248 & 5.749782211793$\cdot 10^{-8}$ & 0.79916095 & 7.901106307$\cdot 10^{-7}$ \\

 ${}^{10}$Be & 4 & 2.3610 & 0.99957389838248 & 5.749782211793$\cdot 10^{-8}$ & 0.70209175 & 6.941535113$\cdot 10^{-7}$ \\
      \hline
 ${}^{10}$B  & 5 & 2.4278 & 0.99933413638122 & 7.219245621776$\cdot 10^{-8}$ & 0.74241787 & 1.862654830$\cdot 10^{-6}$ \\

 ${}^{11}$B  & 5 & 2.4059 & 0.99933413638122 & 7.219245621776$\cdot 10^{-8}$ & 0.72909310 & 1.829251689$\cdot 10^{-6}$ \\
       \hline
 ${}^{12}$C & 6 & 2.4073 & 0.99904101579314 & 8.7092766851788$\cdot 10^{-8}$ & 0.73000871 & 3.905950028$\cdot 10^{-6}$ \\

 ${}^{13}$C & 6 & 2.4614 & 0.99904101579314 & 8.7092766851788$\cdot 10^{-8}$ & 0.76315629 & 4.083351420$\cdot 10^{-6}$ \\

 ${}^{14}$C & 6 & 2.5037 & 0.99904101579314 & 8.7092766851788$\cdot 10^{-8}$ & 0.78958608 & 4.225805319$\cdot 10^{-6}$ \\
     \hline\hline
  \end{tabular}
  \end{center}
  \end{table}
\begin{table}[tbp]
   \caption{The total energies $E$ of some isotope-substituted two-electron ions (in atomic units). $K$ is the total number 
            of basis functions used.}
     \begin{center}
     \begin{tabular}{| c | c | c |}
      \hline\hline
  $K$ & $E({}^{6}$Li$^{+}$) & $E({}^{7}$Li$^{+}$) \\
     \hline\hline
 3500 & -7.279223 0161006 727790 650057 & -7.279321 519787 537196 699113 \\ 
 
 3700 & -7.279223 0161006 727790 650265 & -7.279321 519787 537196 699372 \\ 

 3840 & -7.279223 0161006 727790 650368 & -7.279321 519787 537196 699475 \\

 4000 & -7.279223 0161006 727790 650468 & -7.279321 519787 537196 699574 \\
     \hline\hline
  $K$ & $E({}^{9}$Be$^{2+}$) & $E({}^{10}$Be$^{2+}$) \\
      \hline\hline
 3500 & -13.654709 268248 818671 527237 & -13.654794 978228 935476 431692 \\ 

 3700 & -13.654709 268248 818671 527625 & -13.654794 978228 935476 435740 \\ 

 3840 & -13.654709 268248 818671 527817 & -13.654794 978228 935476 437497 \\

 4000 & -13.654709 268248 818671 528033 & -13.654794 978228 935476 439660 \\
     \hline\hline
  $K$ & $E({}^{10}$B$^{3+}$) & $E({}^{11}$B$^{3+}$) \\
      \hline\hline
 3500 & -22.097340 260098 130926 358406 & -22.098460 503032 611369 090170 \\ 

 3700 & -22.097340 260098 130926 360874 & -22.098460 503032 611369 092636 \\ 

 3840 & -22.097340 260098 130926 362680 & -22.098460 503032 611369 094443 \\ 

 4000 & -22.097340 260098 130926 364302 & -22.098460 503032 611369 096065 \\ 
      \hline\hline
  $K$ & $E({}^{12}$C$^{4+}$) & $E({}^{13}$C$^{4+}$) \\
      \hline\hline
 3500 & -32.404733 488926 278502 692842 & -32.404850 266198 080817 671544 \\

 3700 & -32.404733 488926 278502 693005 & -32.404850 266198 080817 671707 \\

 3840 & -32.404733 488926 278502 693093 & -32.404850 266198 080817 671795 \\

 4000 & -32.404733 488926 278502 693224 & -32.404850 266198 080817 671923 \\
     \hline\hline
  $K$ & $E({}^{14}$C$^{4+}$) & ----- \\
      \hline\hline
 3500 & -32.404949 988753 619902 032262 & ---------------- \\

 3700 & -32.404949 988753 619902 032424 & ---------------- \\

 3840 & -32.404949 988753 619902 032513 & ---------------- \\

 4000 & -32.404949 988753 619902 032644 & ---------------- \\
     \hline\hline
  \end{tabular}
  \end{center}
  \end{table}
 \begin{table}[tbp]
   \caption{The expectation values of the delta-functions and other operators used in calculations of the 
            $\Delta E^{QED}_{\infty}$ and $\Delta E^{QED}_{M}$ corrections for the ground $1^1S(L = 0)-$state in 
            the model two-electron ions with the infinite nuclear masses (in atomic units).}
     \begin{center}
     \begin{tabular}{| c | c | c | c | c |}
      \hline\hline
                                        &  ${}^{\infty}$Li$^{+}$  &  ${}^{\infty}$Be$^{2+}$  &  ${}^{\infty}$B$^{3+}$  &  ${}^{\infty}$C$^{4+}$  \\
                                               \hline\hline
 $\langle \delta({\bf r}_{eN}) \rangle$ & 6.8520 094343 462 & 17.1981 72544 635 & 34.758 743660 935 & 61.443 578056 543 \\

 $\langle \delta({\bf r}_{ee}) \rangle$ & 0.5337 225365 611 & 1.52289 53514 918 & 3.3124 421128 343 & 6.1410 439710 717 \\

 ln $K_0$                               &  5.1798 4912 9  &  5.7550 9181 3  &  6.2014 6720 1  &  6.5662 3588 3  \\

 $\langle (r^{-3}_{ee})_R \rangle$      & -6.5281 0829 296 & -26.725 9651 077 & -70.595 6634 154 & -148.72 6462 987 \\                              
     \hline\hline
  \end{tabular}
  \end{center}
  \end{table}
  \begin{table}[tbp]
   \caption{The expectation values which are needed to determine the mass-dependent components of the isotopic shifts and recoil correction 
            to the lowest-order QED corrections $\Delta E^{QED}_{M}$ (all values are in atomic units) for each isotope.}
     \begin{center}
     \begin{tabular}{| c | c | c | c |}
      \hline\hline
 isotope & $\frac12 \langle {\bf p}^2_1 \rangle$ &  $\langle {\bf p}_1 \cdot {\bf p}_2 \rangle$ &  $\frac12 \langle {\bf p}^2_N \rangle$ \\
     \hline\hline 
 ${}^{6}$Li  & 3.63926634776862256908 & 0.2886900796744086288 & 7.56722277521165376698 \\

 ${}^{7}$Li  & 3.63936484138499574890 & 0.2887308379661048866 & 7.56746052073609638437 \\
       \hline
 ${}^{9}$Be  & 6.82692618052016588002 & 0.4201657546934667785 & 14.0740181157337985385 \\

 ${}^{10}$Be & 6.82701188451767303607 & 0.4202012115590292674 & 14.0742249805943753395 \\
      \hline
 ${}^{10}$B  & 11.0142482768039452882 & 0.5522411301146439512 & 22.5807376837225345276 \\

 ${}^{11}$B  & 11.0143602940252942064 & 0.5522874275672885044 & 22.5810080156178769173 \\
       \hline
 ${}^{12}$C  & 16.2016102297131475001 & 0.6847099993208290105 & 33.0879304587471240106 \\

 ${}^{13}$C  & 16.2017270007914486371 & 0.6847582175983323092 & 33.0882122191812295834 \\

 ${}^{14}$C  & 16.2018267184515755466 & 0.6847993943186503851 & 33.0884528312218014784 \\
     \hline\hline
 isotope & $\langle \delta_{eN} \rangle$ &  $\langle (r^{-3}_{eN})_R \rangle$ & $\Delta E^{QED}_{M}$ $a.u.$ ($\Delta E^{QED}_{M}$ $MHz$) \\
     \hline\hline 
 ${}^{6}$Li  & 6.8501121960089 & -238.6352250738 & 1.1021361193$\cdot 10^{-4}$ (7.2517073026$\cdot 10^{5}$) \\

 ${}^{7}$Li  & 6.8503828685062 & -238.6457736206 & 1.1021844969$\cdot 10^{-4}$ (7.2520256117$\cdot 10^{5}$) \\
       \hline
 ${}^{9}$Be  & 17.195002927213 & -662.1327513745 & 3.3029131600$\cdot 10^{-4}$ (2.1736555344$\cdot 10^{6}$) \\

 ${}^{10}$Be & 17.195319920709 & -662.1462731118 & 3.3029805117$\cdot 10^{-4}$ (2.1732567764$\cdot 10^{6}$) \\
      \hline
 ${}^{10}$B  & 34.752987732681 & -1436.920035269 & 7.5798489928$\cdot 10^{-4}$ (4.9873010540$\cdot 10^{6}$) \\

 ${}^{11}$B  & 34.753508739203 & -1436.943743086 & 7.5799739787$\cdot 10^{-4}$ (4.9873832907$\cdot 10^{6}$) \\
       \hline
 ${}^{12}$C  & 61.435098004106 & -2682.390222121 & 1.4741983427$\cdot 10^{-3}$ (9.6997591317$\cdot 10^{6}$) \\

 ${}^{13}$C  & 61.435752441957 & -2682.421520508 & 1.4742155075$\cdot 10^{-3}$ (9.6998720706$\cdot 10^{6}$) \\

 ${}^{14}$C  & 61.436311305958 & -2682.448248113 & 1.4742301655$\cdot 10^{-3}$ (9.6999685151$\cdot 10^{6}$) \\
     \hline\hline
  \end{tabular}
  \end{center}
  \end{table}
\end{document}